\documentclass[aps,prl,twocolumn,showpacs,superscriptaddress]{revtex4-1}

\usepackage{graphicx}

\newcommand{\ket}[1]{|#1\rangle}

\begin{document}


\title{Four-Photon Quantum Interferometry at a Telecom Wavelength}


\author{Masahiro Yabuno}
\email[]{yabuno@quantum.riec.tohoku.ac.jp}
\address{Research Institute of Electrical Communication, Tohoku University, 
Sendai 980-8577, Japan}

\author{Ryosuke Shimizu}
\address{Center for Frontier Science and Engineering, University of Electro-Communications, Tokyo 182-8585, Japan}

\author{Yasuyoshi Mitsumori}
\address{Research Institute of Electrical Communication, Tohoku University, 
Sendai 980-8577, Japan}

\author{Hideo Kosaka}
\address{Research Institute of Electrical Communication, Tohoku University, 
Sendai 980-8577, Japan}

\author{Keiichi Edamatsu}
\address{Research Institute of Electrical Communication, Tohoku University, 
Sendai 980-8577, Japan}



\date{\today}

\begin{abstract}
We report the experimental demonstration of four-photon quantum interference using telecom-wavelength photons. 
Realization of multi-photon quantum interference is essential to linear optics quantum information processing and measurement-based quantum computing. 
We have developed a source that efficiently emits photon pairs in a pure spectrotemporal mode at a telecom wavelength region, and have demonstrated the quantum interference exhibiting the reduced fringe intervals that correspond to the reduced de Broglie wavelength of up to the four photon `NOON' state. 
Our result should open a path to practical quantum information processing using telecom-wavelength photons.
\end{abstract}

\pacs{03.67.Bg, 42.50.Dv, 42.50.St, 42.65.Lm}

\maketitle

A variety of novel quantum optical technologies have been proposed for use in quantum information processing and quantum metrology \cite{nielsen0521635039, Nature390_575}. 
Photons are the most promising and practical media to demonstrate such novel quantum technologies.
Indeed, linear optical quantum computing \cite{Nature409_46} and measurement-based quantum computing \cite{PhysRevA68_022312} have attracted much attention. 
However, use of the latest quantum technologies requires a large number of multiple photons at the same time.
Furthermore, the photons must be indistinguishable from each other 
because the quantum operations rely on quantum interference between photons.
Thus, we need practical and efficient sources that can provide many, indistinguishable photons.
To date, quantum processing including up to six photons has been demonstrated \cite{NaturePhys3_91, PhysRevLett103_20504}.
These demonstrations used near infrared (800-nm band) photons,
for which efficient photon sources and reliable single-photon detectors are available.  
However, use of telecom-band (1.5-$\mu$m band) photons is desired for practical purposes.
Here we report an experiment with four-photon quantum interference using telecom-wavelength photons, 
in which the photons exhibited reduced ($\propto N^{-1}$) fringe intervals corresponding to the number ($N$) of photons \cite{PhysRevLett82_2868, PhysRevLett89_213601, Nature429_161, Nature429_158, PhysRevA74_33812, Science316_726, Science328_879}.
Combined with the recent developments of novel photon-detecting devices \cite{NaturePhoton3_12},
our result should open a path to practical quantum information processing using telecom-wavelength photons.

The most popular photon sources so far used for the demonstration of quantum information processing are based on spontaneous parametric down-conversion (SPDC),
which generates 
twin (signal and idler) photons 
that can be used as entangled photons \cite{PhysRevLett75_4337} or heralded single photons \cite{OptComm246_545}.
The linear optical quantum computing \cite{Nature409_46} and measurement-based quantum computing \cite{PhysRevA68_022312} 
both rely on quantum interference between photons to carry out quantum operations and quantum measurements.
Photons generated by SPDC must be indistinguishable from each other to make them interfere. 
To do so, spectrotemporal purity of the photons is essential \cite{PhysRevA56_1627}. 
However, in general, photons generated by SPDC have spectrotemporal correlation \cite{OptExpress17_16385} that destroys the  purity of each photon. 
Spectral filtering has often been used to purify the spectrotemporal modes of photons;
however, such filtering inevitably reduces the generation efficiency. 
Thus, efficient generation of spectrotemporal-correlation-free photons is indispensable for multi-photon quantum interference. 

The recent development of group-velocity-matched parametric down-conversion (GVM-SPDC) 
has made it possible to generate spectrotemporal-correlation-free photons with high efficiency \cite{PhysRevA64_063815, NJPhysics10_093011, PhysRevLett100_133601, OptExpress18_3708, PhysRevLett105_253601, PhysRevLett106_013603, PhysRevLett94_083601}. 
In the telecom wavelength region, GVM-SPDC with periodically-poled $\rm KTiOPO_4$ (PPKTP) has been demonstrated \cite{PhysRevLett105_253601, PhysRevLett106_013603, PhysRevLett94_083601, OptExpress17_16385},
and two-photon interference between the generated twin photons has been examined.
For further demonstration of multi-photon interference, one can generate spectrotemporal-correlation-free photons by controlling the pump bandwidth and the phase-matching bandwidth \cite{PhysRevLett105_253601, PhysRevLett106_013603, PhysRevLett94_083601}. 
Furthermore, in the GVM-SPDC, one can use longer crystals than those used in standard SPDC,
and thereby realize the efficient generation of spectrotemporal-correlation-free photons.
Thus, the GVM-SPDC with PPKTP is a promising source for multi-photon quantum interferometry in the telecom wavelength region.

Using a photon source based on the GVM-SPDC, 
we examined a multi-photon (up to four photons) quantum interference that exhibited fringe intervals inversely proportional to the number of photons concerned.
The reduced fringe intervals correspond to the photonic ``de Broglie wavelength" \cite{PhysRevLett74_4835};
an ensemble of multiple photons exhibits the effective de Broglie wavelength of $\lambda/N$, 
where $\lambda$ and $N$  are the classical wavelength of light and the number of the constituent photons, respectively. 
This unique phenomenon can be applied to, for instance, super-resolution \cite{PhysRevLett104_123602} and super-sensitivity \cite{Science316_726} in the phase measurement beyond the classical or the standard quantum limit (SQL).  
It is known that the reduced fringe interval is observed in the multi-photon entangled state called the ``NOON" state 
$(\ket{N, 0}_{\rm AB} - \ket{0, N}_{\rm AB}) / \sqrt{2}$, 
where 
$\ket{N, 0}_{\rm AB}$ $\left(\ket{0, N}_{\rm AB}\right)$
represents the state in which there are $N$ (zero) photons in the optical mode A with zero ($N$) photons in the mode B. 
Thus far, the reduced de Broglie wavelength with the NOON state has been demonstrated for $N=2$ \cite{PhysRevLett82_2868, PhysRevLett89_213601}, $3$ \cite{Nature429_161}, 
$4$ \cite{Nature429_158, PhysRevA74_33812, Science316_726} and $5$ \cite{Science328_879}. 
Note that all these demonstrations were carried out using 800-nm-band photons, 
mainly because efficient photon sources and reliable single-photon detectors are available in this wavelength region.
Our experiment is, to the best of our knowledge, the first demonstration of the four photon quantum interference using telecom band photons. 
\begin{figure}[t]
\includegraphics[width=0.7\hsize ]{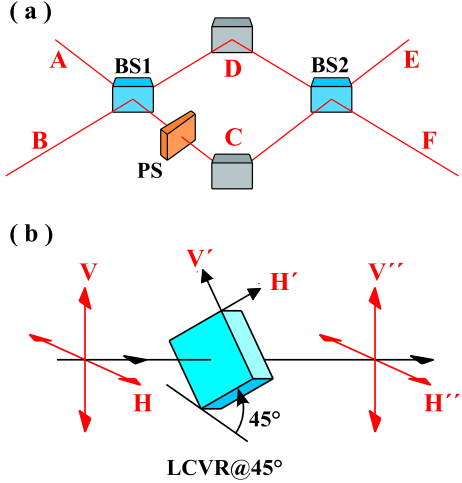}
\caption{\label{fig:MZI} (a) Path-mode and (b) polarization-mode Mach-Zehnder interferometer. 
BS1, BS2: beam splitters; PS: phase shifter; LCVR: liquid crystal variable retarder.
}
\end{figure}

In the following, we explain the principle of our quantum interferometry 
using the conventional Mach-Zehnder interferometer (MZI) as shown in Fig.~\ref{fig:MZI}\,(a).
In the actual experiment, as described later, we used the polarization-mode MZI as shown in Fig.~\ref{fig:MZI}\,(b).
First, we consider the input quantum state $\ket{1, 0}_{\rm AB}$, which contains a single photon in the path mode A and a vacuum in the mode B. 
After BS1 and PS, the state becomes the single photon NOON state: 
$(\ket{1, 0}_{\rm CD} - e^{i\phi}\ket{0, 1}_{\rm CD}) / \sqrt{2}$. 
After BS2, 
the probability of detecting the photon in the mode F is $(1 -\cos\phi)/2$,
which exhibits the normal fringe interval as in the classical interference.
Next, we consider the state $\ket{1, 1}_{\rm AB}$ as the input.
Twin photons generated from degenerate SPDC can be used as this state. 
After BS1 and PS, this state becomes the two photon NOON state: 
$(\ket{2, 0}_{\rm CD} - e^{2i\phi}\ket{0, 2}_{\rm CD}) / \sqrt{2}$. 
Note that the state $\ket{1, 1}_{\rm CD}$ is missing because of the destructive Hong-Ou-Mandel (HOM) interference \cite{PhysRevLett59_2044} at BS1, and that the phase difference is enhanced by a factor of two because the two photons experience the same phase shift together.  
As a result,  the probability of detecting the two photons together in the mode E (or F) is $(1 - {\rm cos}2\phi)/4$, which exhibits a fringe interval that is reduced by a factor of 2 \cite{PhysRevLett89_213601}.
Finally, we consider the state $\ket{2, 2}_{\rm AB}$ as the input.
This state can be generated from a couple of simultaneously generated twin photons from degenerate SPDC, provided that all photons are indistinguishable from each other.
After BS1 and PS,
the state becomes the NOON-like state: 
$\sqrt{3/8}(\ket{4, 0}_{\rm CD} + \ket{0, 4}_{\rm CD}) - \ket{2, 2}_{\rm CD}/ 2$.
This state contains the four photon NOON state as well as the unwanted $\ket{2, 2}_{\rm CD}$ term.
However, since the unwanted term does not produce the state $\ket{3, 1}_{\rm EF}$ after BS2 while the NOON state does,
one can observe the interference that originates only from the NOON state by the post selective detection of the final state $\ket{3, 1}_{\rm EF}$.
The detection probability of the state  $\ket{3, 1}_{\rm EF}$ is $3(1 - {\rm cos}4\phi) / 16$; 
we can observe a fourfold reduction of the fringe interval \cite{PhysRevA65_33820}. 

\begin{figure}[t]
\includegraphics[width=0.7\hsize ]{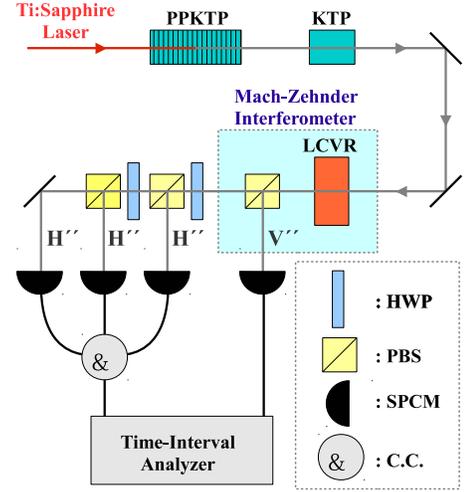}
\caption{\label{fig:Experimental_setup} Experimental layout. LCVR: liquid crystal variable retarder; HWP: half-wave plate; PBS: polarizing beam splitter; SPCM: fiber-coupled single-photon counting module; C.C.: coincidence counter.}
\end{figure}

In the experiment,
we used a polarization-mode MZI with a liquid crystal variable retarder (LCVR) as shown in 
Fig.~\ref{fig:MZI}\,(b).
The function of the polarization-mode MZI is the same as that of the path-mode MZI;
the polarization modes H, V, $\rm H'$, $\rm V'$, $\rm H''$ and $\rm V''$ in Fig.~\ref{fig:MZI}\,(b) correspond to the path modes A, B, C, D, E and F in Fig.~\ref{fig:MZI}\,(a), respectively.
The LCVR whose slow and fast axes ($\rm H'$ and $\rm V'$) are rotated by $45^\circ$ with respect to the original H and V polarizations functions as the beamsplitters (BS1 and BS2) as well as the phase shifter (PS) in the path-mode MZI.  
The polarization-mode MZI provides superior phase stability during the long-time measurement required in our multi-photon experiment.
Figure \ref{fig:Experimental_setup} sketches the experimental setup. 
A mode-locked Ti:sapphire laser operating at a center wavelength of 792 nm and a repetition rate of 80 MHz was used as a pump source of the SPDC. 
A 30-mm-long PPKTP crystal with a poling period of 46.1~$\rm\mu m$ was used for the type-II GVM-SPDC.
The pump light was focused (beam waist:  50~$\rm\mu m$) onto the center of the PPKTP crystal  
and generated twin photons having orthogonal polarizations at a center wavelength of 1584~nm. 
The GVM-SPDC generates the twin photons with a positive spectral correlation \cite{OptExpress17_16385}.
Combined with the negative spectral correlation originating from the pump spectral bandwidth, 
one can control the joint spectral distribution of the produced twin photons \cite{PhysRevLett105_253601, PhysRevLett106_013603}. 
In our experiment, the pump bandwidth (full width at half maximum) was adjusted to be 0.4~nm (corresponding to a pulse duration of 2.3~ps) to generate the photons with eventually no spectrotemporal correlation. 
The joint spectral distribution was observed by a couple of tunable bandpass filters followed by the coincidence detection of the twin photons (not shown in Fig.~\ref{fig:Experimental_setup} ). 
A 15-mm-long KTP crystal whose crystallographic axes are rotated by 90$^\circ$ with respect to the PPKTP crystal compensates for the temporal retardation difference between photons having horizontal (H) and vertical (V) polarizations, 
and thus makes the photons indistinguishable except for their polarizations.
\begin{figure}[t]
\includegraphics[width=0.7\hsize ]{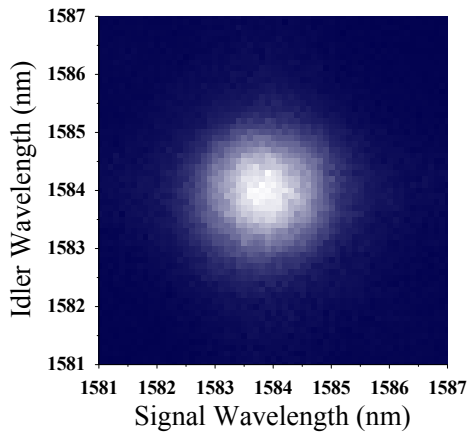}
\caption{\label{fig:Exp2photonspec} Measured joint spectral distribution of photon pairs generated by GVM-SPDC with a PPKTP crystal.}
\end{figure}

Then the photons having H and V polarizations were led into the polarization-mode MZI.
After passing through the MZI, 
the photons were 
led into single mode fibers and then 
detected by the four single-photon detectors (id Quantique id201), 
one of which was for the $\rm V''$ polarization mode and the other three of which were for the $\rm H''$ mode.
The generated twin photon state in the lowest order was $\ket{1, 1}_{\rm HV}$.
For the single-photon interference,
we prepared the state $\ket{1, 0}_{\rm HV}$ by blocking the V-polarized photon with a polarizing beam splitter,
and observed the photon in the $\rm V''$ mode.
For the two-photon interference,
we used the state $\ket{1, 1}_{\rm HV}$ as the input
and observed the photons in the state $\ket{2, 0}_{\rm H''V''}$ by the twofold coincidence between the two detectors in the $\rm H''$ mode.
For the four-photon interference,
the state $\ket{2, 2}_{\rm HV}$, a pair of simultaneously generated twin photons, was used.
This was done by selecting the events with fourfold coincidence detection.
We observed the photons in the final state $\ket{3, 1}_{\rm H''V''}$  by the fourfold coincidence between the three detectors in the $\rm H''$ mode and one detector in the $\rm V''$ mode.

\begin{figure}[t]
\includegraphics[width=0.75\hsize ]{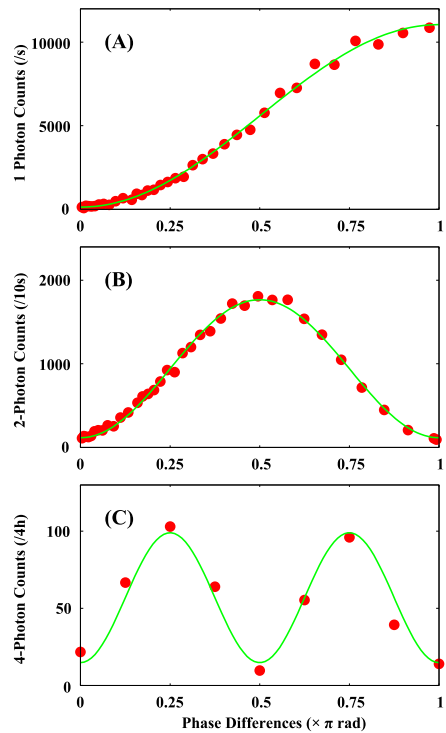}
\caption{\label{fig:4photon_interference} Measured interference patterns on (A) one-, (B) two- and (C) four-photon interference.}
\end{figure}

Figure \ref{fig:Exp2photonspec} shows the measured joint spectral distribution of the photon pairs generated from our photon source. 
The signal and idler photons have almost identical spectral shapes with a bandwidth of 1.7~nm centered at 1583.9~nm.
More importantly, the joint spectral distribution exhibits practically no spectral correlation between the signal and idler,
indicating that the photons are generated in a pure, indistinguishable spectrotemporal mode. 
In fact, the Schmidt number \cite{NJPhysics10_093011} obtained from the measured joint spectral distribution was $K_{\rm exp}$=1.01, indicating almost no spectral correlation. 
This value was in good agreement with the expected Schmidt number ($K_{\rm calc}$=1.01) for the joint spectral distribution $| \sigma(\omega_s,\omega_i) |^2 = | f(\omega_s,\omega_i) g(\omega_s+\omega_i) |^2$ calculated from the amplitudes of phase-matching function  $f(\omega_s,\omega_i)$ and pump spectrum $g(\omega_p)= g(\omega_s+\omega_i) $, where $\omega_p$, $\omega_s$, and $\omega_i$ are the frequencies of pump, signal, and idler photons, respectively.
The Schmidt number expected for the joint spectral amplitude $\sigma(\omega_s,\omega_i) =  f(\omega_s,\omega_i) g(\omega_s+\omega_i)$ was $K$=1.21, 
which corresponds to a good spectral purity of signal or idler photons, i.e, $P$=$K^{-1}$=0.83.
Thus, this source allows us to conduct the multi-photon interference experiment without spectral filtering. 
The photon pair production rate of our source was as high as 48,000 pairs/mW/sec.
In the following experiments, we used a pump power of $\sim$50 mW, 
from which we expect the mean photon-pair number of $3.0\times10^{-2}$ pairs/pulse.
Detailed characteristics of our photon source will be discussed elsewhere. 

Figure \ref{fig:4photon_interference} shows the measured interference fringes for (a) one-, (b) two- and (c) four-photon NOON states, as a function of the single-photon phase difference from 0 to $\pi$. 
We see that
the fringe interval of the $N$-photon interference was $2\pi/N$, exhibiting the $1/N$ reduction expected from the theory.
Also note that the fringe visibilities ($V$) are quite high:
$V$=0.98, 0.88 and 0.74 for one-, two-, and four-photon interferences, respectively.
Our four-photon interference clearly exhibited the reduced fringe interval with high visibility that demonstrated the super resolution beyond the classical limit ($V$=0.20) \cite{PhysRevLett104_123602}.
However, it did not reach the threshold ($V$=0.816) \cite{Science316_726} required to beat the SQL in the phase sensitivity.
The degraded fringe visibility is in part attributable to the imperfect purity ($P$=0.83) of our photon pair source.
The rest of the degradation
might have originated from uncorrelated background counts and the slight spectral difference between the signal and idler photons.

In conclusion, 
we have generated spectrotemporal-correlation-free photons via GVM-SPDC at telecom wavelength.
Using the intrinsically indistinguishable photons thus obtained,
we demonstrated the quantum multi-photon interference up to the four-photon NOON state. 
To our knowledge, this is the first demonstration of four-photon quantum interference using photons all in a telecom wavelength.
To date, the major problems with photonic quantum information processing in the telecom wavelength region were the lack of efficient photon sources and photon detectors.
Recently, we have seen steep progress in photon-detecting devices such as transition edge sensors and superconducting nanowire single photon detectors working in the telecom wavelength region \cite{NaturePhoton3_12}.
Our results,
combined with the latest advancements in novel photon-detecting devices,
will open a path to practical quantum information processing using telecom-wavelength photons.

\begin{acknowledgments}
This work was supported by a Grant-in-Aid for Creative Scientific Research (17GS1204) and a Grant-in-Aid for JSPS Fellows (22$\cdot$7182) from the Japan Society for the Promotion of Science, and JST PRESTO program. 
\end{acknowledgments}


\end{document}